# Coexistence of ferroelectricity and superconductivity in a two-dimensional monolayer

Jianyong Chen[1], Wen-Yi Tong[4], Ping Cui[2,3,], and Zhenyu Zhang[2,3,]

[1]*Institute of Quantum Materials and Physics, Henan Academy of Sciences, Zhengzhou, 450046, China*

[2]*International Center for Quantum Design of Functional Materials (ICQD),Hefei National Research Center for Physical Sciences at Microscale, University of Science and Technology of China, Hefei, Anhui 230026, China*

[3]*Hefei National Laboratory, University of Science and Technology of China, Hefei 230088, China*

[4]*Key Laboratory of Polar Materials and Devices (MOE) and Department of Electronics, East China Normal University, Shanghai, 200241 China*

The coupling of ferroelectricity (FE) and superconductivity (SC) becomes the frontier of condensed matter research recently especially in the realm of two-dimensional (2D) materials. Identifying a general strategy to realize coexistence of FE and SC in a single material is extremely important for this active field, but quite challenging thus far. We show in this work that coexistence of robust FE and metallicity/SC can be realized by hole-doping a ferroelectric insulator which hosts antibonding highest valence bands (HVB). Using typical 2D ferroelecrtic SnS monolayer as a concrete example, we demonstrate that 0.30 hole/cell doping leads to enhancement of total polarization mainly ascribed to the increasing of polar displacement and ionic polarization. In addition, due to the strong Fermi surface nesting and prominent softening of out-of-plane acoustic phonon upon hole-doping, SnS can be turned into a single gap superconductor with an unexpectedly high transition temperature ($T_c$) of ~7 K, whereas the polar phonon mode gives negligible contribution to electron-phonon couplings. Our work provides general principle and realistic material for realizing metallic FE and superconducting FE, which paves the way for reversible and nonvolatile superconducting devices.

## I. INTRODUCTION

Recently, the interplay between FE and SC has become a hot topic among both theoretical and experimental researchers [1-3]. On one hand, the physical mechanism including quantum critical point induced or enhanced SC is attracting [4-6], on the other hand, ferroelectric field control of SC pave the way for new type nanoscale high-density superconducting applications including reversible topological SC [7-10], reversible superconducting diode effect [11], cryogenic memory elements [12], and gate-tunable qubits [13].

Research on the coupling between FE and SC can be dated back to 1999, where FE-modulated SC was observed in a heterostructures consisting of ferroelectric $Pb(Zr_xTi_{1-x})O_3$ and superconducting $GdBa_2Cu_3O_{7-x}$ [14]. In 2006, a complete switching between zero-resistance SC and the normal state was realized by ferroelectric polarization reversal using Nb-doped $SrTiO_3$ as the superconducting channel and ferroelectric $Pb(Zr,Ti)O_3$ as the gate oxide [15]. The most prominent modulation of $T_c$ of 30 K was achieved latter [16]. The emergence of 2D ferroelectrics [17-22], 2D superconductors [23-25] and their van der Waals (vdW) stackings empower more diverse and exotic tunabilities. Recently, ferroelectric reversed superconducting diode effect in Cu intercalated bilayer $NbSe_2$ [11], ferroelectric tuning of band topology and SC in $IrTe_2/In_2Se_3$ heterobilayer [7,8], ferroelectric topological superconductors [9] and ferroelectric topological nodal-point superconductors [10] in 2D FEs proximated with superconductors have been theoretically proposed.

The aforementioned couplings of FE and SC based on ferroelectric/superconductor heterostructures are external and suffers from interfacial imperfections. Realizing the coexistence FE and metallicity (or SC) in a single material is physically intriguing and appealing for new technologies, but commonly viewed to be incompatible due to the screening of long-range Coulombic interaction between dipoles by itinerant electrons. It was suggested by Anderson and Blount 50 years ago that ferroelectric metal can occur if the iterant electrons interact weakly with the ferroelectric phonons [26]. Although seemingly be simple, it was not until 2013 that the first unambiguous intrinsic polar (irreversible) metal $LiOsO_3$ was reported experimentally [27]. Coexistence of metallicity and "ferroelectric-like" state was reported in several doped 3D ferroelectrics including $SrTiO_3$ [28], $BaTiO_3$ [29] and $HfO_2$ [30] from experimental signature or theoretical calculations. Despite extensive studies, the general connection between chemical bonding and polarization in a metal remains elusive. What's more, due to the non-layered bonding and 3D metallic character, the reversal of polarization by an electric field, i.e., ferroelectricity, is prohibited due to the screening of external electric field and indeed has not been experimentally confirmed in either $LiOsO_3$ [27] or $SrTiO_3$ [28].

The first truly switchable out-of-plane polarization in a metal was reported in bi- and tri-layered $WTe_2$ [31] (and bulk-like $WTe_2$ [32]). Later, coexistence/coupling of FE and SC were observed clearly in magic-angle twisted bilayer graphene (MATBG) with aligned boron nitride layers [1] and bilayer $MoTe_2$ [2]. The polarizations in those systems are limited to the out-of-plane direction and other ferroelectric quantum states (e.g. quantum paraelectric) are inaccessible. In principle, the appearance of polarization and its reversibility in MATBG, $WTe_2$ and $MoTe_2$ rely on the interlayer vdW coupling and their atomic thickness. The vdW coupling is chemically strong enough to induce unequal distribution of charge along the out-of-plane direction, and physically weak enough to forbid the free flow of carriers between adjacent layers. The atomic thickness allows the penetration of external electric field. Going beyond this scope and investigating the coexistence of FE and metallicity/SC in a monolayer (not vdW stacked bilayer) would not only deepen our understanding of the fundamental physics of FE but also vastly enrich this active field.

To this end, revisiting the driving force of FE is imperative. Ferroelectric transition is determined by a competition between two energies. These are the short-range Coulomb repulsions between the valence electron clouds around the ions, which favor the nonpolar phase, and the long-range dipole–dipole Coulomb interactions, which favor the FE phase. Since free carriers screen dipole–dipole interactions, metallicity is commonly believed to be detrimental to FE [33,34]. In general, the polarization in a ferroelectric comes from both the electronic and ionic contributions; the electronic polarization is associated with the charge ordering of multiple valences [35,36], while the ionic polarization describes the inconsistency between the positive and negative ionic centers. In terms of ionic polarization, the most well-known driving force is the pseudo Jahn–Teller effect [37], where the spontaneous electron transition (added covalency) drives high-symmetry lattice to a low-symmetry lattice, i.e. ferroelectric distortion [34].

In this work, we show that, in contrast to conventional wisdom that charge doping is detrimental to FE, hole-doping an insulating ferroelectric with antibonding HVB (2D SnS as a concrete example) results in robust enhancement of total polarization by increasing the ionic part contribution, via the strengthening of covalent bonding. Based on accurate first-principles calculations, we show that hole-doping in an experimental accessible concentration induced EPC is strong enough to trigger SC with a $T_c$ of ~7 K. HVBs can be an effective descriptor to search more potential doped ferroelectric superconductors. This paper is organized as follows. Sec. II describes the computational details for various quantities. Sec. III presents the main findings of this work including the evolution of polarization with both electron and hole-doping and SC in hole-doped SnS. In Sec. IV, we give a discussion concerning superconducting diode effect and others. A conclusion is given in Sec. V.

**II. COMPUTATIONAL DETAILS**

Calculations related with ferroelectric switching pathway and band topology are performed using the Vienna Ab initio Simulation Package (VASP) [38,39]. The generalized gradient approximation (GGA) in the Perdew–Burke–Ernzerhof (PBE) [40] exchange-correlation functional and the projector augmented wave (PAW) [41] formalism were applied. The kinetic energy cut-off was set at 500 eV, and the Brillouin zone was sampled by Γ-centered $17 \times 17 \times 1$ k points using the Monkhorst–Pack scheme [42]. The convergence threshold for self-consistent-field iteration was set to be $10^{-6}$ eV and the atomic positions were fully relaxed until the forces on each atom were less than 0.01 eV/Å. The ferroelectric switching pathway is calculated by using the climbing image nudged elastic band (CINEB) method [43]. The bonding nature associated with various electron energy bands was analyzed using the COHP method. The COHP method partitions the energy of the band structure into interactions between pairs of atomic orbitals between adjacent atoms. It is a bond-weighted measure of the electronic density of states (DOS) and provides a quantitative measure of the bonding and antibonding contributions to the band energy. Importantly, for this work, the sign of the COHP differs for bonds with bonding or anti-bonding nature: a positive (negative) sign corresponds to anti-bonding (bonding) interactions. By convention, COHP diagrams plot the negative value (-pCOHP) such that bonding (anti-bonding) states on the right (left) of the axis can be easily visualized. We further quantify the degree of anti-bonding using the integrated area under the COHP curves with respect to the electron band energy. The spontaneous polarization value polarizations are calculated by the Born effective charge using $\delta P = (e/\Omega)\sum_i Z_i^* d_i$, where $e$ is the electronic charge, $\Omega$ the unit cell volume, $Z_i^*$ the Born effective charge (BEC) of atom $i$, and $d_i$ its relative displacement in the polar direction. This formula has been widely used to simulate polarization of metals or doped insulators [44-46].

The orbital selective external potential (OSEP) method [47] was used to investigate the influence of orbital hybridization. A projector operator $|inlm\sigma\rangle\langle inlm\sigma|$ that allows the external potential $V_{ext}$ to influence the specific atomic orbital $|inlm\sigma\rangle$ was introduced. The original Kohn-Sham Hamiltonian changes to $H^{OSEP} = H^0 + |inlm\sigma\rangle\langle inlm\sigma| V_{ext}$. Using this method, the hybridization strength between two atomic orbitals can be easily modified, which is very helpful for analyzing the orbital effect [44,48].

Electron-phonon coupling (EPC) calculations were performed using Quantum-ESPRESSO [49] package. The Perdew-Burke-Ernzerh (PBE) of parameterized generalized-gradient approximation (GGA) [40] was used to describe the exchange correlation, unless specified otherwise. Norm-conserving pseudopotentials were adopted to represent the interaction between the ionic cores and valence electrons. A plane-wave basis with a 120 Ry energy cutoff was used to represent electronic wave functions. Carrier doping was simulated with a jellium model, where excess or defect electronic charge was compensated by a uniform neutralizing background. The dynamical matrices and EPC were calculated using the density functional perturbation theory (DFPT) [50] in the linear response regime. Electronic self-consistent calculations were performed with a $40 \times 40 \times 1$ **k**-point grids with a Marzari-Vanderbilt [51] smearing of 0.02 Ry. The dynamical matrix was computed on a $8 \times 8 \times 1$ Monkhorst–Pack (MP) [52] **q**-point grids.

The $\lambda$ can be calculated via the isotropic momentum-independent Eliashberg function $\alpha^2 F(\omega)$ [53]:

$$\lambda = 2\int_0^\infty d\omega \alpha^2 F(\omega)/\omega = \sum_{\mathbf{q}v} \lambda_{\mathbf{q}v}, \tag{1}$$

$$\alpha^2 F(\omega) = \frac{1}{2\pi N(\varepsilon_F)} \sum_{\mathbf{q}v} \frac{\gamma_{\mathbf{q}v}}{\omega_{\mathbf{q}v}} \delta(\omega - \omega_{\mathbf{q}v}), \tag{2}$$

where $\lambda_{\mathbf{q}v}$ is the branch ($v$)- and momentum (**q**)-resolved EPC strength, $N(\varepsilon_F)$ is the density of states at the Fermi level, $\varepsilon_F$ is the corresponding energy at the Fermi level, and $\gamma_{\mathbf{q}v}$ and $\omega_{\mathbf{q}v}$ are the phonon linewidth and phonon frequency of the phonon branch index $v$ with the wave vector **q**, respectively. The $\lambda_{\mathbf{q}v}$ is defined as

$$\lambda_{\mathbf{q}v} = \frac{2}{\hbar N(\varepsilon_F)} \sum_{\mathbf{k}jj'} \left| g_{\mathbf{k}+\mathbf{q}j',\mathbf{k}j}^{\mathbf{q}v} \right|^2 \delta(\varepsilon_{\mathbf{k}j} - \varepsilon_F) \delta(\varepsilon_{\mathbf{k}+\mathbf{q}j'} - \varepsilon_F) / \omega_{\mathbf{q}v}, \tag{3}$$

the EPC matrix element $g_{\mathbf{k}+\mathbf{q}j',\mathbf{k}j}^{\mathbf{q}v}$ represents the probability of electron scattering from an initial electron state j with momentum **k** (the corresponding single-particle energy $\varepsilon_{\mathbf{k}j}$) to a final electron state $j'$ with **k**+**q** (the corresponding single-particle energy $\varepsilon_{\mathbf{k}+\mathbf{q}j'}$), mediated by a phonon with momentum **q** and branch index $v$ (the corresponding frequency $\omega_{\mathbf{q}v}$). The phonon linewidth stems from the Fermi's golden rule:

$$\gamma_{\mathbf{q}v} = 2\pi\omega_{\mathbf{q}v} \sum_{\mathbf{k}jj'} \left| g_{\mathbf{k}+\mathbf{q}j',\mathbf{k}j}^{\mathbf{q}v} \right|^2 \delta(\varepsilon_{\mathbf{k}j} - \varepsilon_F) \delta(\varepsilon_{\mathbf{k}+\mathbf{q}j'} - \varepsilon_F). \tag{4}$$

According to Eqs. (3) and (4), $\lambda_{\mathbf{q}v} = \gamma_{\mathbf{q}v} / (\pi \hbar N(\varepsilon_F) \omega_{\mathbf{q}v}^2)$. The cumulative EPC is given by

$$\lambda(\omega) = 2 \int_0^\omega d\omega' \alpha^2 F(\omega') / \omega'. \tag{5}$$

The $T_c$ is determined by the McMillian-Allen-Dynes formula [54,55]:

$$T_c = \frac{\omega_{\log}}{1.2} \exp\left[\frac{-1.04(1+\lambda)}{\lambda - \mu^*(1+0.62\lambda)}\right], \tag{6}$$

$$\omega_{\log} = \exp\left[\frac{2}{\lambda} \int_0^{\omega_{\max}} \alpha^2 F(\omega) \frac{\ln(\omega)}{\omega} d\omega\right], \tag{7}$$

where $\omega_{log}$ is the logarithmic average of the phonon frequencies, The $\mu^*$ measures the strength of the electron-electron repulsion, and its value usually lies in the range of 0.10-0.15 [56].

To more precisely capture the superconducting gap and $T_c$, we solve the anisotropical anisotropic Migdal-Eliashberg equations with the EPW package [57,58]. The maximally localized Wannier functions (MLWFs) [59] are constructed on a uniform unshifted $16 \times 16 \times 1$ **k**-point grid; Sn-s, Sn-p and S-p orbitals are chosen as projectors. The interpolated fine **k**- and **q**-point grids are $160 \times 160 \times 1$ and $80 \times 80 \times 1$, respectively. The Matsubara frequency cutoff is set to five times the largest phonon frequency, and the Dirac $\delta$ functions are replaced by Lorentzians of widths 25 and 0.05 meV for electrons and phonons, respectively. Electrons within ±200 meV from the Fermi energy are taken into the EPC process.

## III. RESULTS

In most ferroelectric insulators or even normal insulator, the HVB are composed of bonding states and the conduction bands arises from the antibonding states. Introducing carriers of either electron or hole type will on one hand screen the dipole-dipole interaction, i.e. reduction of electronic polarization, and weaken the covalent bond strength (at least true for ferroelectrics driven by pseudo-Jahn-Teller effect), i.e. reduction of ionic polarization (Fig. 1(a)). Therefore, doping is usually detrimental to the survive of FE. However, there is a class of unconventional insulators, where in momentum-space, the HVBs are composed of antibonding states between the positive and negative charged ions (Fig. 1(b)). In this case, the electron doping destroys the polarization as expected. However, since hole-doping cuts down antibonding states, the total covalent bonding can be increased. This leads to an elongated polar displacement and enhancement of ionic polarization. Finally, the total ferroelectric polarization upon hole doping can be preserved or even surpass the magnitude of the undoped case, depending on the balance between the reduction of electronic polarization and reinforcement of ionic polarization. In short, hole-doping a ferroelectric insulator that hosts antibonding HVB provides a reasonable and feasible route for realizing coexistence of FE and metallicity/SC. From the materials aspect, the HVBs of in-plane polarized SnS (and its family including GeS, SnSe, SnTe) monolayer, out-of-plane polarized $CuInP_2S_6$ monolayer exhibit antibonding character. In contrast, the HVBs manifests as nonbonding and bonding characters for $BaTiO_3$ and $HfO_2$ respectively as shown in Fig. 1(c).

In the following, we verify the above proposal by systematically investigating the polarization of charge-doped 2D ferroelectric SnS via first-principles calculations. SnS is one member of the family of group-IV mono-chalcogenides MX (M=Ge, Sn; X=O, S, Se) [60], in which GeS, GeSe, SnS, and SnSe are predicted to be multiferroic with coupled FE and ferroelasticity [61-63]. We choose SnS as an example, since SnX is more chemically stable than GeX and the Sn-S bonding is stronger [64]. Furthermore, monolayer SnS has been successfully prepared and its room-temperature in-plane FE is confirmed experimentally [65]. The optimized lattice constants ($a$ = 4.08 Å, $b$ = 4.28 Å) and ferroelectric flip barrier (40 meV) agrees well with previous theoretical reports [61,62], indicating the reliability of our calculation. Undoped SnS monolayer exhibits as a direct band gap semiconductor with the valence band maximum near Y point and conduction band minimum near X point in the Brillouin zone [60] (Fig. 2 (a)). The projected density of states (DOS) are shown in Fig. 2(b), the lowest two bands (-7.5 eV ~ -4.0 eV) comes from Sn-s orbitals, which also contribute a large proportion to the two HVBs. The upper six valence bands are dominated by S-p orbitals, and the lowest six unoccupied conduction bands comes from Sn-p orbitals. From the perspective of molecular orbital diagram, we see that the two HVBs are formed by the antibonding states between Sn-s and S-p orbitals. This is confirmed by the COHP calculation where the value in this region is negative (i.e. antibonding character).

First-principles calculations reveal that ionic parts gives rise 82% of the total polarization of SnS, whereas electronic dipole moment contributes only 18%. The ferroelectric distortion in SnS originates from pseudo-Jahn-Teller effect, i.e., enhanced bonding between Sn-p and S-p orbitals after polar displacement [66]. First, we investigate the evolution of bonding strength by calculating the integrated COHP up to the Fermi level (ICOHP). For electron-dopings, the ICOHP of out-of-plane bonds $b_\perp$ (Fig. 3(a) and blue lines in 3(b)) decreases monotonously and significantly with the increasing of doping levels. The ICOHP of in-plane bonds $b_{//}$ (red lines in Fig. 3(b)) also decreases as compared with undoped case but not in a monotonous manner. Due to the overall weakening of covalent bonding, the bond length of $b_{//}$ becomes larger than that of the undoped one (green lines) and the polar displacement between Sn and S in the *y* direction are also shorter (Fig. 2(c)). We calculated the polarizations of the systems under two different conditions (i.e. fixed atomic positions (green lines in Fig. 2(d)) and fully relaxed atomic positions (red lines in Fig. 2(d))) to distinguish the effect of ionic displacements. It is shown in Fig. 2(d) that, the polarizations of electron-doped systems for both conditions are smaller that undoped one, which can be understood by the screening of dipole interaction by free carriers and the reduction of ionic contributions.

The behavior of polarization under hole doping is different as shown in Fig. 3(b). The ICOHP of $b_\perp$ keeps unchanged from that of undoped case in the whole doping levels investigated. For the bonding of $b_{//}$, the ICOHP keeps unchanged when doping level is lower than 0.20 hole/cell. Then, the ICOHP increases gradually with the increasing of doping levels, and the ICOHP surpass the magnitude of the undoped one. The increasing of ICOHP leads to the shortening of the $b_{//}$ bond length for the 0.30 hole/cell doping (Fig. 3(b)). The most prominent effect of hole doping is the increasing of Sn-S displacement along *y* direction, with 9% increment upon 0.30 hole/cell doping (Fig. 3(c)). If the atoms are fixed to the positions of undoped one to investigate pure electronic effect, one can see directly that the polarizations are almost the same with the undoped one (Fig. 3(d)). Taking the ionic changes into effect by fully relaxing the atomic positions, we see that the polarization first decreases (≤0.10 hole/cell) and then increases (≥0.15 hole/cell) with the increasing of hole dopings and the polarization increases by 4% (0.04 $P_0$) at the doping level of 0.30 hole/cell. Note that the increment of polarization (4%) is lower than that of Sn-S displacement along $y$ direction (9%), demonstrating the decreasing of Born effective charges with doping, based on $\delta P = (e/\Omega)\sum_i Z_i^* d_i$. The flip barriers develop a consistent trend with that of polar displacement. The value of barriers ($E_b$) are moderate (20 meV < $E_b$ < 40 meV) and the system still lies safely in the ferroelectric energy valleys.

To verify the crucial role of antibonding valence states in enhancing the polarization under hole doping. We weaken the antibonding states by shifting down the Sn s orbital by 2 eV using the OSEP method [47]. Here Sn-s instead of S-p orbital is selected, because S-p orbital also interacts with Sn-p orbital, which makes the situation complex and unclear. First, we test the influence on undoped SnS monolayer. When atoms are fixed to the positions of the undoped structure, the polarization decreases slightly by 6%. After fully relaxing the atomic positions, the displacement of Sn from S along $y$ shortens obviously by 32% (from 0.318 Å to 0.216 Å). Meanwhile, the polarization based on this structure decreases obviously by 32% as well (red filled star in Fig. 3(d)), reflecting the decisive role of atom displacement arise from orbital interactions in inducing polarization. Next, we see its influence for fixed doping level (0.3 hole/cell doping as a typical example, the atom positions are fixed), the polarization reduces slightly by 4%. After relaxing the atoms, the displacement of Sn from S shortens obviously by 21% (from 0.348 Å to 0.274 Å), and the polarization based on this structure decreases by 14%, and now the polarization of ~0.9 $P_0$ (black filled star in Fig. 3(d)) is much lower than undoped one. Finally, in the unified condition that Sn-s orbital is shifted down by 2 eV, 0.3 hole/cell doping enhances the polarization by 0.21 $P_0$ (difference between black and red stars), much larger than the enhancement 0.04 $P_0$ in orbital-unshifted case. This test not only confirms that antibonding HVB can enhance polarization, but also tell us that tuning the orbital interaction is efficient in enhancing polarization upon hole-doping.

To summarize this section, the central physic of hole-doping enhanced polarization as compared with undoped case is rooted in the antibonding HVBs. Hole-doping leads to strengthening of bonding (increased total ICOHP) via reducing occupied antibonding states, which results in increasing of polar displacements and polarization. The antibonding HVBs are the prerequisite for enhancement of polarization upon hole doping. Whereas the magnitude of enhancement by hole doping is positively related with the strength of antibonding (Sn-s, S-p antibonding in the case of SnS). To re-affirm our proposal, we hole-dope a ferroelectric that has weak bonding or nonbonding HVBs, namely $BaTiO_3$. Results show that polarizations keeps unchanged in the nonbonding region (0.5 eV deep into valence bands), and decreases slightly in the weak bonding region (0.5 eV ~ 1.55 eV deep into the valence bands), which is consistent with previous results [45,67].

Having established the hole-doping enhanced FE, we next investigate the possibility of phonon-mediated SC by first analyzing the electronic properties. At small doping levels (for instance 0.15 hole/cell), the band maxima close to Y first cross the Fermi level. Fig. 4(b) displays the isosurface of electronic states at ±10 meV around the Fermi level, which localize around each atom and prevent the coupling of electrons with atomic vibrations (i.e. phonons). Besides, there are only two small Fermi pockets along the Γ-Y line (Fig. 4(c)) and the DOS at Fermi level is only 0.30 states/spin/eV, indicating a small EPC. The projected band structures of 0.30 hole-doped SnS are shown in Fig. 4(d-g). The band maximum near Y is dominated by Sn-s and S-pxy orbitals, whereas the band maximum near Γ comes from Sn-pz and S-pz states. Due to the participation of large amount of Sn-pz and S-pz states, the states near Fermi level spans in a broader space and fulfil the void space between Sn and S atoms in the $z$ direction, but leaving the in-plane direction un-overlapped. In this case, the Fermi pocket near Y expands and additional Fermi pocket near Γ appears. The existence of multiple Fermi pockets and multi-dimensional (both in-plane and out-of-plane states) electronic states opens more channels for electrons scattering and has profound influence on both phonon dispersion and total EPC (see following analysis). Fig. 4(e,f) display the Fermi surfaces and nesting function of 0.30 hole-doped SnS. There is an oval-shape Fermi pocket centered at Γ and two bowl-like Fermi pockets near Y. It can be seen from Fig. 4(f) that strong nesting occurs at several typical nesting vectors $\mathbf{q}_{nest}$= ±0.10 Y, $\mathbf{q}_{nest}$ = ±0.38 Y, $\mathbf{q}_{nest}$ = ±0.73 Y (depicted in Fig. 4(i)), which will exert direct influence on the phonon frequency and EPC.

Then we turn to the phonon dispersion of SnS monolayer. In three dimensional polar materials, macroscopic electric fields are present in the long-wavelength limit ($\mathbf{q} \to 0$), the polar longitude optical (LO) mode exert additional long-range Coulomb force on atoms and lift up the phonon frequency near the center of Brillioun zone as compare with the transverse optical (TO) mode, resulting in the ubiquitous phenomenon called LO-TO splitting. It is expected that the strong electronic polarization in SnS would bring prominent LO-TO splitting as well. Indeed, by using conventional calculation technique, i.e. modeling a 2D monolayer by using a 3D unit cell with a large vacuum, we see a large LO-TO splitting (red lines in Fig. 5(a)). To be specific, the frequency of TO mode associated with S vibration along *x* direction (TO@*x*) is 110 cm$^{-1}$, but the LO mode (LO@*x*) rises to 221 cm$^{-1}$. Similarly, the frequency of TO mode associated with S vibration along *y* direction (TO@*y*) is 141 cm$^{-1}$, and the LO mode (LO@*y*) rises to 210 cm$^{-1}$. However, Sohier *et al* realize that the analytical model based on Born effective charges and dielectric tensor for 3D is not suitable for 2D materials since the artificial periodic images brings spurious interactions that yields unphysical results in the long-wavelength limit regardless of the thickness of vacuum [68]. By using the LO-TO splitting treatment for real 2D materials in QE package, we found, at variance with 3D expectations that, the LO-TO splitting vanishes completely at the zone center for 2D monolayer, and the LO mode develops a steep slope related with Born effective charge near the zone center (blue lines in Fig. 5(a)). The universal breakdown of LO-TO splitting in 2D materials is experimentally verified in a typical polar material h-BN [69]. Upon hole doping, the system is turned into a metal, since the macroscopic electric field disappears (recognized by the QE package), the behavior of LO modes changes drastically, i.e., the frequencies of LO modes falls to the same value of TO modes at Γ point, and the sharp down-turn close to Γ (blue lines in Fig. 5(a)) is replaced by a gentle downhill to Γ. This indicates that the significant "softening" of the two LO modes is not an indicator of strong EPC appear in normal nonpolar metals, but a consequence of changing of dielectric environment upon charge doping. Apart from the LO-TO modes, the highest two in-phase and out-of-phase optical modes involving with the vibration of S atoms along *z* directions also exhibits obvious softening in the whole Brillioun zone upon hole doping. In addition, other modes display slight softening as well.

Now, we quantitatively investigate the doping-induced EPC. As shown in Fig. 5(b), the highest-two optical branches and the lowest acoustic branches of 0.30 hole/cell doped SnS exhibit strong phonon-softening as compared with the frequency of 0.15 hole/cell doped case which is beneficial for EPC as $\lambda_{\mathbf{q}\nu} \propto 1/\omega_{\mathbf{q}\nu}^2$ Fig. 5(b) presents the phonon spectra of 0.30 hole/cell doped SnS monolayer decorated with the branch- and momentum-resolved EPC constant $\lambda_{\mathbf{q}v}$ where larger disk represent larger $\lambda_{\mathbf{q}v}$. The green lines are nesting function along the high-symmetry lines. It is apparent that the momentum of larger $\lambda_{\mathbf{q}v}$ coincides with the peak of nesting functions (except the largest $\lambda_{\mathbf{q}v}$ at Γ), indicating the crucial role of nesting in enhancing EPC. The Eliashberg function $\alpha^2F(\omega)$ in Fig. 5(c) has two peaks in the low frequency region and two peaks in the high frequency region, whereas the the magnitude of intermediate phonons are negligible. Compare with the case of 0.15 hole/cell doping, the peaks in the high frequency region are lowered and the magnitude is intensified. The broadened peak in the range of 0~40 cm$^{-1}$ is strongly enhanced, and the sharp peak located near 50 cm$^{-1}$ is also strongly intensified and slightly moves to higher frequency. The $N(\varepsilon_F)$ are 0.30 and 0.87 states/spin/eV for 1.50 and 0.30 hole/cell doped cases. The EPC of $\lambda$ 0.15 hole/cell doped SnS is mediate with $\lambda = 0.58$, due to the dominance of low-frequency phonons in EPC, the $T_c$ is estimated to be a small value of 1.48 K ($\mu^* = 0.10$) using the McMillian-Allen-Dynes formula. The total isotropical $\lambda$ amounts to 1.50 for 0.30 hole/cell doping and the $T_c$ is estimated to be 5.35 K ($\mu^* = 0.10$) using the McMillian-Allen-Dynes formula.

By inspecting the evolution of $\lambda$ with frequency, it is clear that the low frequency phonons dominate the total EPC. To be specific, the $\lambda$ from phonons in the range of 0~70 cm$^{-1}$ contribute 84% of the total $\lambda$. Fig. 5(d) gives the phononic momentum space distribution of $\lambda_\mathbf{q}$ for the three acoustic branches (branch 1 to 3) and two optical branches (branch 4 to 5), in which the out-of-plane acoustic branch provides the largest contributions, especially along the Γ-Y direction. Branch 2 and 3 are in-plane acoustic vibrations. Branch 4 is associated with the in-plane vibration of Sn atoms along $x$ axis. Branch 5 involves with mixed out-of-plane and in-plane vibration of both Sn and S atoms. In addition, the vibration direction of Sn and S atoms in the same Sn-S pairs is the same, but opposite for different pairs. The $\lambda_{\mathbf{q}v}$ of branch 5 at Γ displays a exceptionally larger value than others. Overall, the EPC of hole-doped SnS are dominated overwhelmingly by out-of-plane vibrational phonon modes, which is consistent with the dominance of out-of-plane electronic states at Fermi level. The distribution of total $\lambda$ (the rightmost and the lowest panel in Fig. 5(d)) nearly coincides with the shape of nesting function (Fig. 4(j)), again demonstrates the importance of nesting. We note that polar phonon mode at Γ with $\omega = 141$ cm$^{-1}$ (two Sn and two S atoms vibrate oppositely along $y$ axis) which is responsible for the ferroelectric transition possess negligible EPC (nearly vanishing $\alpha^2F(\omega)$). The weak EPC of polar phonon in hole doped SnS resembles the situation in polar metal LiOsO$_3$ [70].

Because the Li atoms dictated polar phonons spatially separate from the conduction electrons from Os and O orbitals. Different from SnS and LiOsO$_3$, the polar modes in doped SrTiO$_3$ [71] and BaTiO$_3$ [29] has strong EPC, with the former originates from quantum paraelectric fluctuation induced coupling and the latter from classical EPC.

In previous sections, we stress the role of Fermi nesting in EPC. It is clear from Eq. (3) that $\lambda_{\mathbf{qv}}$ of a particular **q** is proportional to nesting function described by double delta functions. However, the role of nesting on the total EPC is not so direct. The total $\lambda$ can be decomposed and relates with phonon softening effect and nesting [72] as:

$$\lambda = \mathrm{N}(\varepsilon_F)(V_0 + \frac{2|M_c|^2}{\omega'^2}) , \qquad (8)$$

where $V_0$ are from phonons that are unaffected by nesting, $|M_c|^2$ denotes an effective coupling matrix element of the relevant phonon modes, and $\omega'$ is the renormalized phonon frequency due to nesting. For an acoustic mode, $\omega'^2(\mathbf{q}) = \omega_0^2(\mathbf{q}) - |g_{ep}(\mathbf{q})|^2 \chi_0(\mathbf{q})$ [73], where $\omega_0(\mathbf{q})$ is the bare phonon frequency, $g_{ep}(\mathbf{q})$ and $\chi_0(\mathbf{q})$ are the EPC matrix and particle-hole susceptibility. Since $\chi_0(\mathbf{q})$ diverges logarithmically at the Fermi nesting vectors, relevant phonon will be drastically softened [72,74,75], resulting in an enhancements of $\lambda$. However, if the $g_{ep}(\mathbf{q})$ is nearly zero or even negative, nesting enhanced EPC will be absent. In a word, the prerequisite for nesting enhanced EPC is a nesting induced phonon-softening (the credibility can be verified with EPC results using LDA functional in Fig. 6).

$T_c$ is the most central quantity of SC. The isotropical Eliashberg theory assumes that the EPC of the states on Fermi surface are the same, which is not suitable for the case of hole-doped SnS where the Fermi surface shows obvious anisotropy. To more accurately calculate the EPC and $T_c$, we solve the anisotropical Migdal-Eliashberg equations using the EPW package [57,58]. As expected, the $\lambda$ shows strong anisotropy on the Fermi surface. The distribution of $\lambda_{n\mathbf{k}}$ on the Fermi pocket centered on Γ is quite uniform and moderate. In contrast, the $\lambda_{n\mathbf{k}}$ on the Fermi pocket near Y shows strong anisotropy, the Fermi line parallel with X has the largest $\lambda_{n\mathbf{k}}$ due to perfect nesting of nearby two Fermi lines, whereas other parts of the Fermi pocket near Y has the lowest $\lambda_{n\mathbf{k}}$ because these state are composed mainly by in-plane S-pxy states where the deformation potential or EPC matrix is small. Different from famous two-gap superconductor MgB$_2$ [76], the anisotropical superconducting gap of SnS exhibits a one-gap nature, and the average value is 1.3 meV. The gap decrease gradually with the increasing of temperature and close at ~ 7 K, which is identified as the $T_c$. The value here is larger than that obtained by McMillian-Allen-Dynes formula, which is common in anisotropical systems like doped-graphene [77].

To confirm the strong EPC and the emergence of SC, we recalculate all quantities using a different exchange functional, i.e. LDA functional. As shown in Fig. 6(a), the bandwidth of the highest two valence bands reduces slightly, and the Fermi pocket near Y point also shrinks slightly as compared with results from PBE (dashed lines). Although the overall changes of electronic structures are quite weak, LDA leads to drastic modifications of phonon dispersion and EPC. First, the highest two phonon branches are hardened significantly by ~25 cm$^{-1}$ with LDA. Second, the prominent phonon softening of lowest acoustic branches near the Γ point is significantly reduced, which demonstrates that those phonon-softening are not determined solely by Fermi surface nesting (since the Fermi nesting in LDA is nearly unchanged), but in a combined effect with local chemical coupling of electrons and phonons (Eq. (8)). On the other side, the significant hardening of acoustic phonons at LDA level indicate that hole doped SnS is not on the verge of dynamical instability like charge density wave or structural broken. Due to the overall hardening of phonon frequencies, the EPC is reduced and the total $\lambda$ reduces to 0.88. But the most significant contribution of EPC are still from 0~50 cm$^{-1}$, consistent with previous analysis based on PBE. The estimated $T_c$ based on Allen-Dynes formula is 4.4 K for LDA. We further calculate the anisotropical superconducting gaps as shown in Fig. 6(d). The superconducting gap at 1 K for LDA is slightly smaller than that of PBE, but still observable. Going beyond the framework of phonon-mediated SC, the strong Fermi nesting in hole-doped SnS is a direct signature of spontaneous Fermi surface instability which may also lead to pure electronic-driven SC [78,79].

At the end of this part, we list three aspects supporting the feasibility of our predictions. First, previous experiments demonstrate that the crystal structure of hole-doped few-layer SnS (not single layer) remains stable and the ferroelectric polarizations are enhanced by hole doping, which strongly support part of our predictions [80]. Second, the maximum doping concentration considered in this work equals to $1.7 \times 10^{14}$ cm$^{-2}$ (0.30 hole/cell). This is experimentally accessible considering that doping concentration as high as $3.5 \times 10^{15}$ cm$^{-2}$ has been achieved [81]. Besides gating, proper substrate (for instance SnSe$_2$ [82]) can naturally realize hole doping. Finally, we believe that the SC can be observed by various experimental groups as the $T_c$ of SnS lies above the temperature of liquid helium (~4 K).

## IV. DISCUSSION

Before closing, we discuss some potential implications based our findings. Lots of efforts have been paid to studying diode effects in intrinsic superconductors and Josephson junctions, wherein the critical supercurrent differs in two opposite directions. On one hand, researchers became aware that the interpretation of experimental data is often complicated by extrinsic factors related to Joule heating, vortex dynamics and device geometry (including asymmetry of surfaces, asymmetric layer structures, and the asymmetric pinning centers), which all affect the asymmetry of the critical current and diode efficiency. On the other hand, one major challenge in the creation of Josephson junctions made from multi-materials are the engineering of ultra-clean interfaces between the different material species, which is needed for efficient coupling between different phases [83]. We propose to solve this dilemma based our predictions. Using the hole-doped SnS region as superconducting electrode and the undoped region as tunnel barrier, a gate-defined superconducting diode can be obtained. Most interestingly, the polarity of diode can be easily reversed by flipping the direction of ferroelectric polarization.

The first observation of superconducting diode effect without an applied magnetic field or magnetic material is reported recently in 2022. The Josephson junction is based on an inversion symmetry breaking van der Waals heterostructure of $NbSe_2/Nb_3Br_8/NbSe_2$. It is proposed that the polarization embedded in the heterostructure is crucial for field free superconducting diode effect [84,85]. However, the underlying mechanism is not settled down so far partly due to the ubiquitous device asymmetry. Considering that tunnel barrier made by SnS (or other in-plane polarized ferroelectrics) can be tuned to nonpolar or polar continuously by external gating without introducing extra chemical or physical changes, gated-defined Josephson junctions (hole-doped SnS as SC electrode, undoped or electron-doped SnS as tunnel barrier) may serve as a testbed for clarifying the relationship between superconducting diode effect and polar atomic structures.

The coexistence of FE and SC in hole-doped SnS is also attracting for its diverse tunabilities by both doping level and strains. For instance, flip barrier decreases dramatically upon uniaxial compressive strain along $y$, which are 6.5 meV and 2 meV for 2% and 3% strains. On the basis of this high tunability, different quantum regimes can be accessed in one single material: (i) If the barrier is too small, quantum fluctuations prevent the developing of macroscopic polarization due to the tunneling between the 'up' and 'down' polar states, and finally the material develops quantum paraelectric behavior. (ii) If the barrier is in proximity with the QCP and the ferroelectric-like order is ionic displacement induced, the soft mode fluctuations provide the pairing interaction for SC carriers and exhibits cooperation between ferroelectric-like order and SC as exemplified in doped $SrTiO_3$ [5,28,71]. (iii) If the barrier is far above the ferroelectric quantum critical point, stable FE holds, which provides two bistable states with opposite internal polarizations.

Finally, due to the existence of superconducting gaps, the screening of polarization exists in normal metals may be significantly suppressed in superconductors, and the incompatibility of FE and SC may be reconciled to a large extent. In addition, nodeless superconductors have superconducting quasiparticle gap at every momentum points, thus the polarization based on Berry phase is well defined in terms of the combined normal electrons and superconducting quasiparticles. Further theoretical and experimental studies are needed to clarify those issues.

## V. Conclusion

In summary, we propose that the existence of antibonding HVBs in an insulating ferroelectric enables unreduced or even enhanced polarization upon hole doping, which is ascribed to the enhanced atomic distortion originated form pseudo-Jahn-Teller effect. We confirm this proposal in hole-doped SnS monolayer based on first-principles calculations. In addition, the unique strongly nested Fermi surface and doping-induced phonon softening in hole-doped SnS leads to strong electron-phonon coupling and emergence of SC. The $T_c$ determined by solving anisotropic Migdal-Eliashberg equations with PBE functional is 7.2 K for 0.30 hole/cell doped SnS. Our work provides a general strategy for realizing polar metal or metallic ferroelectrics, and provides a realistic material for investigating coupling of FE and SC both on the mechanism and device applications.

## ACKNOWLEDGMENTS

This work was supported by the Innovation Program for Quantum Science and Technology (Grant No. 2021ZD0302800), the National Natural Science Foundation of China (Grants No.

11974323 and No. 12204121), the National Key R&D Program of China (Grant No. 2017YFA0303500), the Anhui Initiative in Quantum Information Technologies (Grant No. AHY170000), and the Strategic Priority Research Program of Chinese Academy of Sciences (Grant No. XDB30000000). J.C. also acknowledges the support from the Guangxi Natural Science Foundation (Grants No. 2019GXNSFBA245077 and No. 2021GXNSFAA220129).

**Figures and Figure captions**

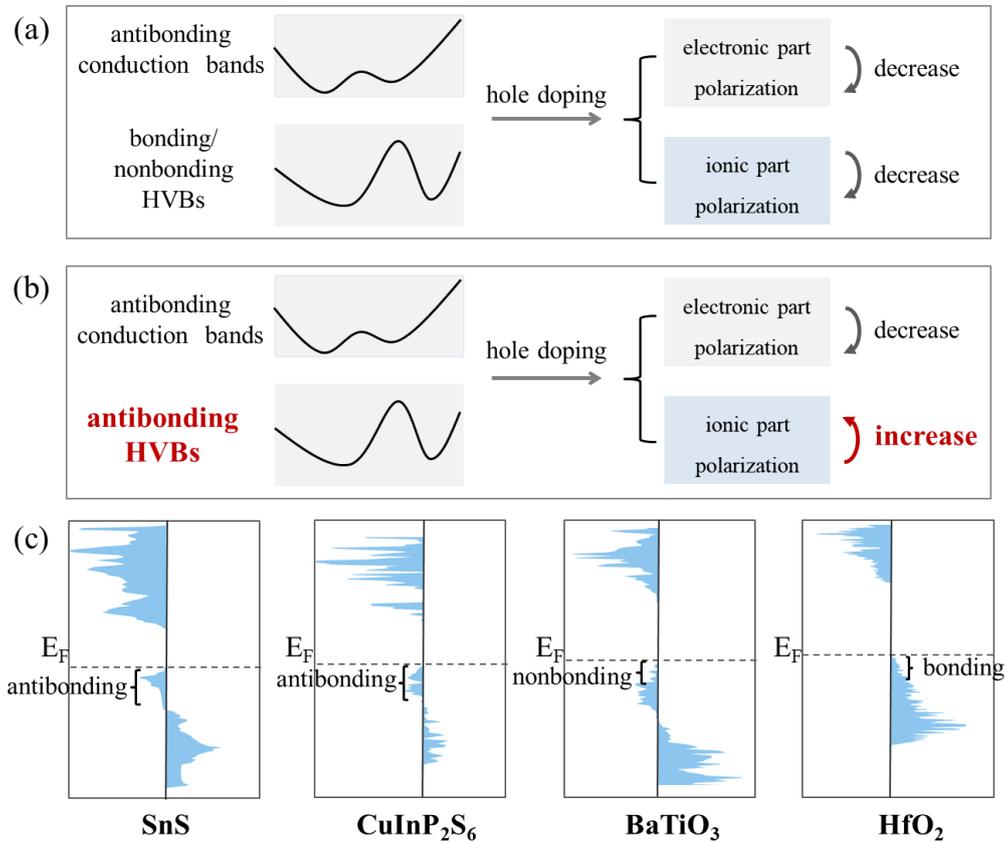

**FIG. 1.** Route towards coexistence of metallicity and ferroelectricity by hole-doping an insulating ferroelectric with antibonding HVBs. Evolutions of polarization by doping two different kinds of insulating ferroelectrics: (a) bonding/nonbonding HVBs and (b) antibonding HVBs. (c) First-principle calculated -COHP of typical 3D and 2D ferroelectrics. The HVBs of pure in-plane polarized SnS monolayer and pure out-of-plane polarized $CuInP_2S_6$ monolayer exhibit prominent antibonding character. The HVB of $BaTiO_3$ shows nonbonding character and $HfO_2$ shows pure bonding character. All Fermi levels are set at the top of valence bands.

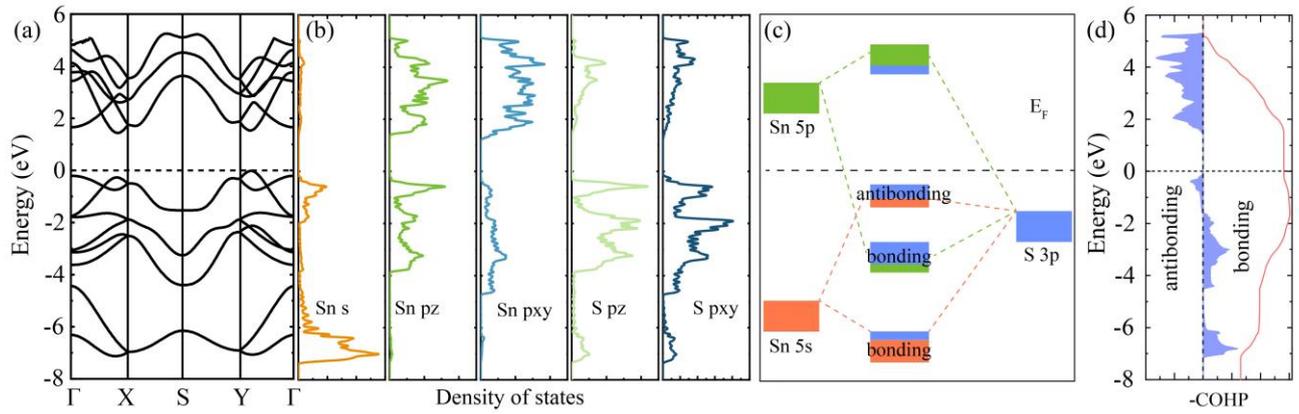

**FIG. 2.** (a) Band structure of undoped SnS monolayer along a typical high symmetry path. (b) Atom and orbital projected density of states (DOS). (c) Molecular orbital diagram of SnS. Note that the interaction between Sn-5s and S-3p generates two occupied anti-bonding HVBs. (d) -COHP (shaded blue) and ICOHP (red line) diagram of SnS, confirming the strong anti-bonding character of the two HVBs.

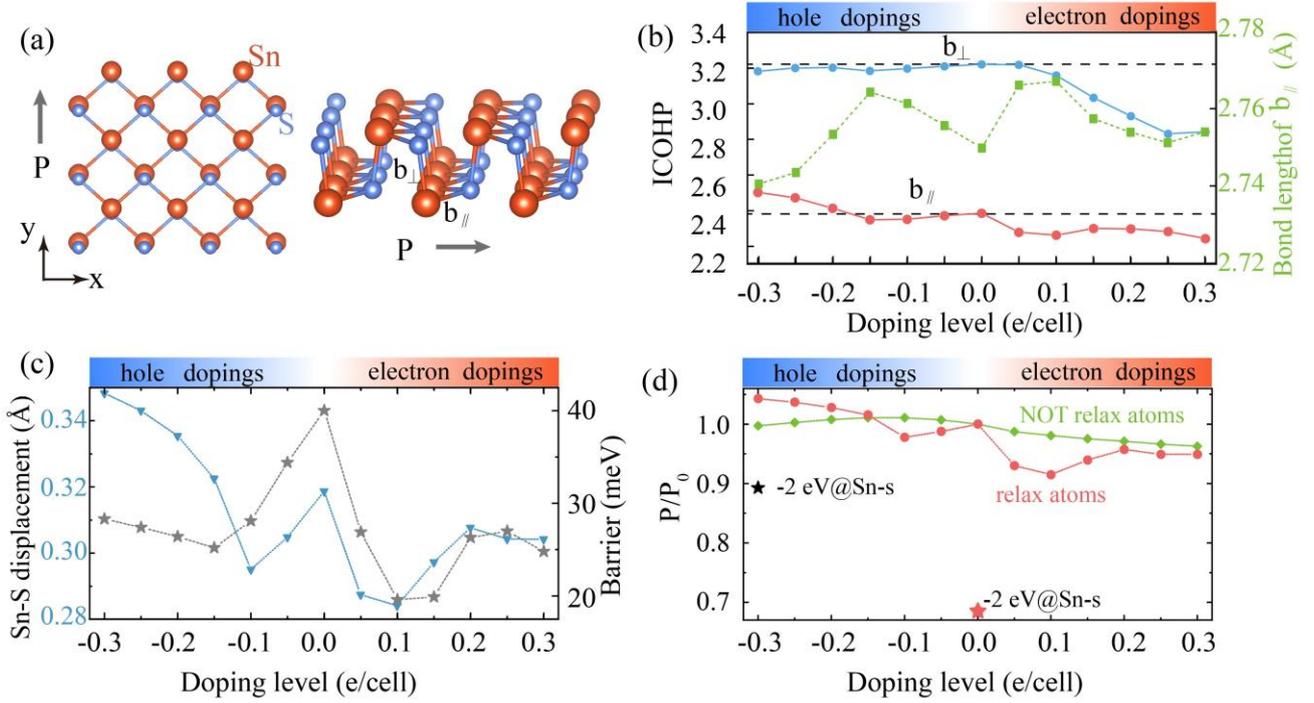

**FIG. 3.** (a) Top and side view of the SnS monolayer, the in-plane and out-of-plane directed bonds are denoted as $b_{\parallel}$ and $b_{\perp}$. (b) Evolution of ICOHP of $b_{\parallel}$ and $b_{\perp}$ (left) and bond length of $b_{\parallel}$ (right) with doping levels. (c) Evolution of Sn-S displacement along *y* axis and NEB barriers with doping levels. (d) Evolution of magnitudes of polarization with doping levels. The red (green) dots are results from structures with atom positions fully relaxed (unchanged from the undoped one). The red and black stars are polarizations for undoped and 0.30 hole/cell doped systems calculated by dereasing the Sn-s orbital by 2 eV using OSEP method and fully relaxing the atomic positions. All the polarization magnitudes are renormalized to the value of undoped SnS.

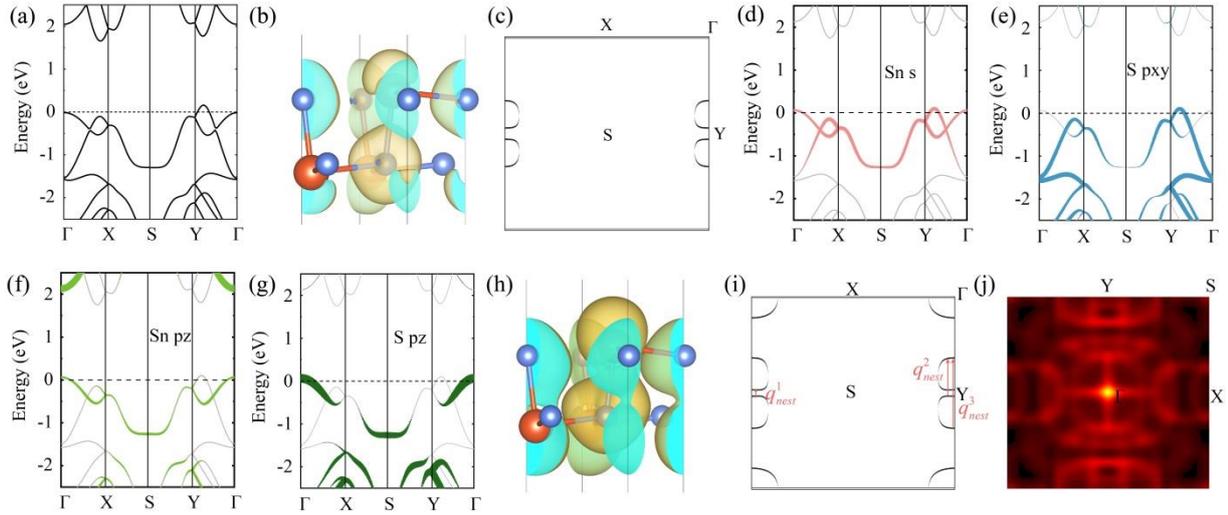

**FIG. 4.** (a-c) Electronic properties of 0.15 hole/cell doped SnS. (a) Band structure, (b) Isosurface of electronic states at ±10 meV around the Fermi level. (c) Fermi surfaces. (d-j) Electronic properties of 0.30 hole/cell doped SnS. (d-g) Atom and orbital projected band structures, (h) Isosurface of electronic states at ±10 meV around the Fermi level. (i) Fermi surfaces, (j) Fermi nesting function.

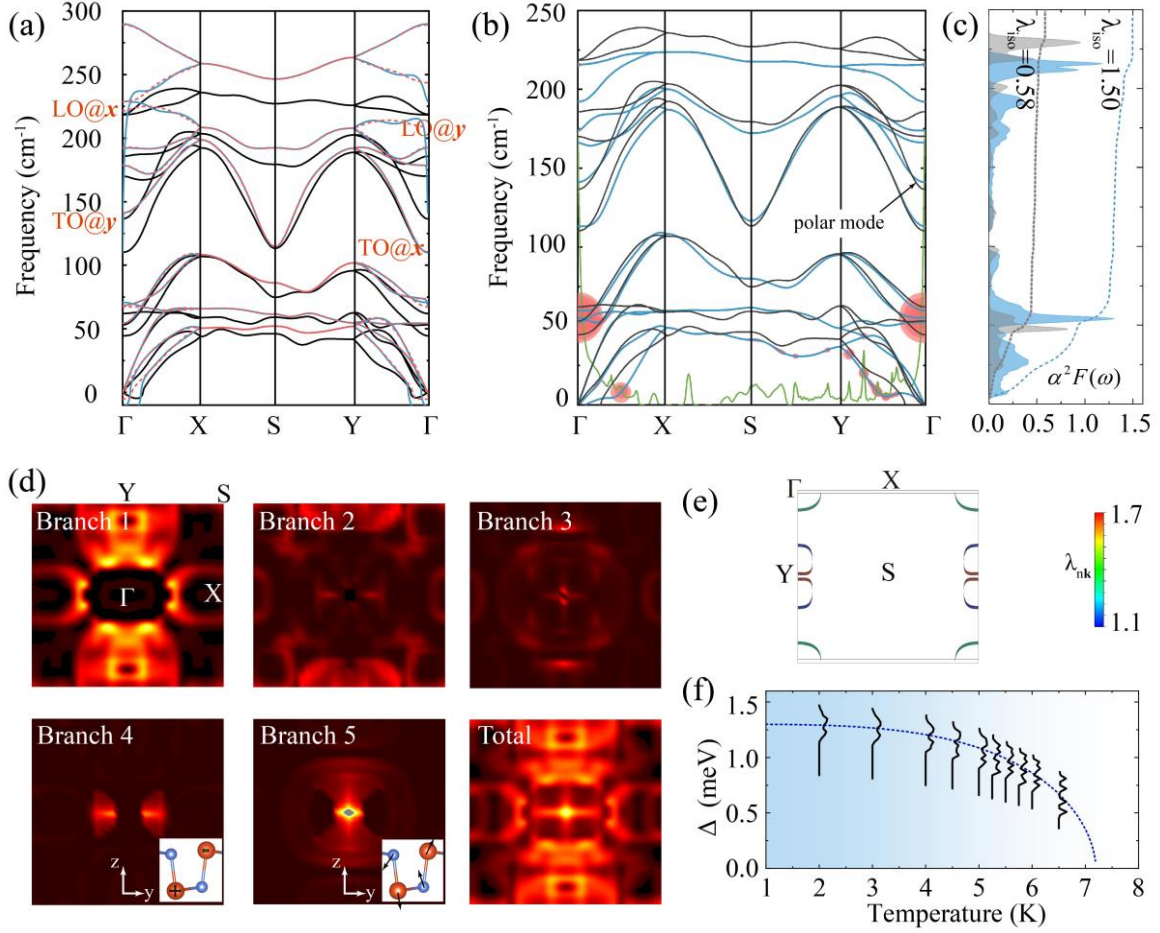

**FIG. 5.** (a) Phonon spectra of SnS monolayer. Red dashed (blue) lines for undoped SnS using 3D (2D) LO-TO splitting treatment. Black lines for 0.15 hole/cell doped SnS. (b) Phonon spectra of 0.15 (gray) and 0.30 (blue) hole/cell doped SnS. The radii of red disks are proportional to the branch- and momentum-resolved $\lambda_{\mathbf{q}\nu}$ of 0.30 hole/cell doped SnS. Green line displays the Fermi nesting function along the given high-symmetry path. The arrow points to the polar mode. (c) Eliashberg spectral function $\alpha^2F(\omega)$ (shaded region) for 0.15 (gray) and 0.30 hole/cell (blue) doped SnS. The dashed lines represent the cumulative frequency-dependent integrated EPC $\lambda(\omega)$. (d) Phononic momentum (**q** space) distribution of $\lambda_{\mathbf{q}\nu}$ for the lowest five branches and total $\lambda$ of 0.30 hole/cell doped SnS. The vibration pattern at $\Gamma$ point for Branch 4 and 5 are shown in the insets. (e) Band and electronic-momentum (**k** space) resolved $\lambda_{n\mathbf{k}}$ for electrons on the Fermi surface of 0.30 hole/cell doped SnS, hoter color indicates larger magnitude. (f) Evolution of the anisotropic superconducting gap as a function of temperature for 0.30 hole/cell doped SnS. The dashed line is a guide for the eyes.

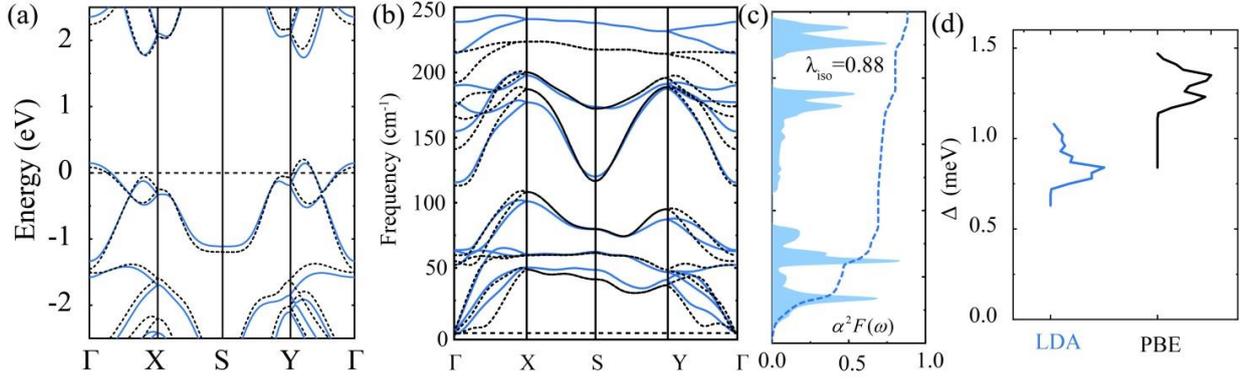

**FIG. 6.** (a) Band structures, (b) phonon spectra (c) Eliashberg spectral function and integrated EPC, (d) evolution of superconducting band gap with temperatures for 0.3 hole/cell doped SnS calculated with LDA functional (blue lines). The black lines in (a), (b) and (d) are results obtained with PBE functional for comparison.